\documentclass[12pt,a4paper]{article}
\usepackage[margin=2cm]{geometry}
\usepackage[english]{babel}
\usepackage[latin1]{inputenc}
\usepackage{amsmath, amsthm, amssymb}
\usepackage[pdftex]{graphicx}
\usepackage{amssymb}
\usepackage{latexsym}
\usepackage{amstext}
\usepackage{epstopdf}
\usepackage{floatflt}

\newcommand{\aci}{\mathbf{a_5}}
\newcommand{\ad}{\mathbf{a_2}}

\newcommand{\adc}{\mathbf{a_2^c}}
\newcommand{\add}{\mathbf{a_2^{(2)}}}
\newcommand{\adf}{\mathbf{a_2^f}}
\newcommand{\adgd}{\mathbf{a_2^{(2)}}}
\newcommand{\adgu}{\mathbf{a_2^{(1)}}}

\newcommand{\adu}{\mathbf{a_2^{(1)}}}

\newcommand{\alg}{\textbf{\large $\mathfrak A$}}
\newcommand{\au}{a_1}
\newcommand{\aud}{a_1^\dagger}
\newcommand{\aun}{\mathbf{a_1}}
\newcommand{\aut}{a_1^T}

\newcommand{\bu}{b_1}
\newcommand{\bud}{b_1^\dagger}

\newcommand{\but}{b_1^T}
\newcommand{\be}{\begin{equation}}
\newcommand{\bea}{\begin{equation}\begin{array}}
\newcommand{\beas}{\begin{equation*}\begin{array}}
\newcommand{\bef}{\begin{flalign}}
\newcommand{\befs}{\begin{flalign*}}
\newcommand{\bes}{\begin{equation*}}
\newcommand{\cc}{\mathbb{C}}
\newcommand{\ctre}{\mathbf{c_3}}
\newcommand{\cu}{c_1}

\newcommand{\cut}{c_1^T}
\newcommand{\ee}{\end{equation}}
\newcommand{\eea}{\end{array}\end{equation}}
\newcommand{\eeas}{\end{array}\end{equation*}}
\newcommand{\eef}{\end{flalign}}
\newcommand{\eefs}{\end{flalign*}}
\newcommand{\ees}{\end{equation*}}
\newcommand{\ekm}{\varepsilon_k^-}

\newcommand{\ekmp}{\varepsilon_k^{\mp}}
\newcommand{\ekp}{\varepsilon_k^+}

\newcommand{\ekpm}{\varepsilon_k^{\pm}}

\newcommand{\ejpm}{\varepsilon_j^\pm}
\newcommand{\eo}{\mathbf{e_8}}

\newcommand{\es}{\mathbf{e_6}}
\newcommand{\est}{\mathbf{e_7}}

\newcommand{\fq}{\mathbf{f_4}}
\newcommand{\fqe}{\mathfrak{f}}
\newcommand{\gII}{\textbf{\large ${\mathfrak g}_{II}$}}
\newcommand{\gIII}{\textbf{\large ${\mathfrak g}_{III}$}}

\newcommand{\gd}{\mathbf{g_2}}

\newcommand{\gon}{\mathbf{g_0^n}}
\newcommand{\gzd}{\mathbf{g_0^2}}
\newcommand{\gzo}{\mathbf{g_0^8}}
\newcommand{\hu}{\mathbf{H}}

\newcommand{\ingr}[2]{\includegraphics[scale={#1}]{#2}}

\newcommand{\jdot}{\!\cdot\!}
\newcommand{\jo}{\mathbf{J}}

\newcommand{\jotn}{\mathbf{J_3^n}}
\newcommand{\jobtn}{\mathbf{\overline J_3^{\raisebox{-2 pt}{\scriptsize \textbf n}}}}
\newcommand{\jotu}{\mathbf{J_3^1}}
\newcommand{\jobtu}{\mathbf{\overline J_3^{\raisebox{-2 pt}{\scriptsize \textbf 1}}}}
\newcommand{\jotd}{\mathbf{J_3^2}}
\newcommand{\jobtd}{\mathbf{\overline J_3^{\raisebox{-2 pt}{\scriptsize \textbf 2}}}}

\newcommand{\joto}{\mathbf{J_3^8}}
\newcommand{\jobto}{\mathbf{\overline J_3^{\raisebox{-2 pt}{\scriptsize \textbf 8}}}}

\newcommand{\lk}{\mathfrak{L}}

\newcommand{\Lvxp}{L_{\mathbf x^+}}
\newcommand{\Lvxm}{L_{\mathbf x^-}}
\newcommand{\Lvxpm}{L_{\mathbf x^\pm}}

\newcommand{\Lvyp}{L_{\mathbf y^+}}
\newcommand{\Lvym}{L_{\mathbf y^-}}
\newcommand{\Lvypm}{L_{\mathbf y^\pm}}

\newcommand{\mep}{\star}

\newcommand{\nbf}{\mathbf{n}}
\newcommand{\nin}{\noindent}
\newcommand{\oo}{\textbf{\large $\mathfrak C$}}

\newcommand{\qq}{\mathbb{Q}}
\newcommand{\rep}{\mathbf \varrho}

\newcommand{\rmp}{\rho^\mp}

\newcommand{\rpm}{\rho^\pm}
\newcommand{\rr}{\mathbb{R}}

\newcommand{\str}{\text{str}}
\newcommand{\sut}{\mathbf{su(3)}}

\newcommand{\vx}{\mathbf x}

\newcommand{\vxm}{\mathbf x^-}
\newcommand{\vxp}{\mathbf x^+}
\newcommand{\vxpm}{\mathbf x^\pm}
\newcommand{\vy}{\mathbf y}
\newcommand{\vym}{\mathbf y^-}
\newcommand{\vyp}{\mathbf y^+}
\newcommand{\vypm}{\mathbf y^\pm}

\newcommand{\xs}{x^\#}

\numberwithin{equation}{section}
\begin{document}
%
\begin{titlepage}
\begin{center}

\hfill DFPD/2015/TH/14


\vskip 2cm

{\huge{\bf Exceptional Lie Algebras\\ at the very Foundations \vspace{7pt} \\of Space and Time}}

\vskip 2.5cm

{\large{\bf Alessio Marrani\,$^{1,2}$ and  Piero Truini\,$^3$}}

\vskip 20pt

{\em $^1$ Centro Studi e Ricerche ``Enrico Fermi'',\\
Via Panisperna 89A, I-00184, Roma, Italy \vskip 5pt }

\vskip 10pt

{\em $^2$ Dipartimento di Fisica e Astronomia ``Galileo Galilei'', \\Universit\`a di Padova,\\ Via Marzolo 8, I-35131 Padova, Italy \vskip 5pt }

{email: {\tt Alessio.Marrani@pd.infn.it}} \\

    \vspace{10pt}

{\it ${}^3$ Dipartimento di Fisica, Universit\` a degli Studi\\
via Dodecaneso 33, I-16146 Genova,  Italy}\\\vskip 5pt
\texttt{truini@ge.infn.it}

    \vspace{10pt}

\end{center}

\vskip 2.2cm

\begin{center} {\bf ABSTRACT}\\[3ex]\end{center}
While describing the results of our recent work on exceptional Lie and Jordan algebras,
so tightly intertwined in their connection with elementary particles, we will try to stimulate a critical discussion on the nature of spacetime and indicate how these algebraic structures can inspire a new way of going beyond the current knowledge of fundamental physics.

\vskip 2.2cm

\begin{center}

Talk presented by P.T. at the Conference\\ {\em Group Theory, Probability, and the
Structure of Spacetime} in honor of V.S.Varadarajan,\\ UCLA 
Department of Mathematics, November 7--9, 2014. \\To appear in a special issue of {\em ``p-Adic Numbers, Ultrametric
Analysis and Applications''}.

\end{center}

\end{titlepage}

\newpage \setcounter{page}{1} \numberwithin{equation}{section}

\newpage\tableofcontents

\section{Introduction}

\begin{center}
{\it Life is like a box of chocolates: you never know what you're gonna get.} Forrest Gump
\end{center}

\nin {\it I did not know what I was going to get when I first came to UCLA with my ambitious mission. The mission was to meet the famous professor Varadarajan and invite him to Genoa. The aim was to convince him to start a collaboration with Enrico Beltrametti, Gianni Cassinelli and their students. I knew it would be difficult but I did not hesitate: the goal was definitely worth the risk. Years later Raja told me how close the mission was to fail. But it did not and it has been rewarding far beyond all expectations: it has given birth to both a long-lived collaboration and a lifelong friendship. Raja has given so much to all of us and to me in particular. He has always been there for me every time I needed help.\\ Opening that box of chocolates was a wonderful turn of my life.}\\
Piero\\

This talk is divided into two distinct parts.\\

The first part is based on the results of two papers, \cite{pt1} \cite{T-2} (see also \cite{T-3}), and presents a unifying view of the exceptional Lie algebras through the key notion of {\it Jordan Pairs}.\\

The second part is a digression on fundamental physics and a suggestion on how the mathematical structure presented in the first part may be exploited for a new approach to the many open problems, in particular the problem of unifying gravity with the other interactions of Nature.\\

Before jumping into some inevitable mathematical details on Jordan Pairs and their relationship with the exceptional Lie algebras, we now present an outlook of  these structures, with the hope to convince the reader that they are far less cumbersome than they look at a first glance.\\

Exceptional Lie algebras do indeed appear as fairly complicated objects. The best way to represent them exploits the algebra of the octonions as a bookkeeping device that allows saving a lot of writing in their mathematical realization. Octonions, as it always occurs in Mathematics when you find a useful structure, turn out to have a deeper role than mere bookkeeping.\\

The smallest exceptional group is the automorphism group of the octonions $\oo$, hence its Lie algebra, $\gd$, is the {\it derivation algebra} of the algebra $\oo$. The derivation algebra of an algebra $\alg$ is the set of {\it all} derivations $D$ defined as the subspace of the associative algebra of linear operators on $\alg$ satisfying $D(xy) = D(x)y+xD(y)$, for all $x,y \in \alg$. Since the commutator $[D_1, D_2]$ of two derivations is a derivation of $\alg$, the derivation algebra is a Lie algebra, whose Jacobi identity, in turn, is the {\it Leibniz rule} which defines a derivation.\\

The root diagram of $\gd$ consists of the diagram of its subalgebra $\ad$, the familiar $\sut$ for physicists, plus its $\mathbf{3}$ and $\overline{\mathbf{3}}$ representations.\\

\begin{figure}[htbp]
\begin{center}
\ingr{1}{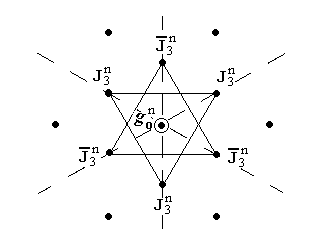}
\end{center}
\caption{A unifying view of the roots of exceptional Lie algebras}
\label{diagram}
\end{figure}

A very similar situation occurs for all other simple exceptional Lie algebras, \cite{pt1}. If we project their roots on the plane of an $\ad$ subdiagram (every such algebra has at least one) we see the same root diagram of $\gd$, plus the roots of a subalgebra, denoted by $\gon$ in figure \ref{diagram}, projected on the center.\\

Nothing surprising so far, but now comes the magic. In place of the $\mathbf{3}$ and $\overline{\mathbf{3}}$ we find three families of {\it Jordan Pairs} $(\jotn, \jobtn)$, each of which lies on an axis,
symmetrically with respect to the center of the diagram. Each pair doubles, through an involution,
a simple Jordan algebra of rank $3$, $\jotn$, the algebra of $3\times 3$
Hermitian matrices over $\hu$, where $\hu=\rr,\,\cc  ,\,\qq,\,\oo$ for $\nbf=1,2,4,8$ respectively, stands for a Hurwitz algebra. The exceptional Lie algebras $\fq$, $\es$, $\est$, $\eo$ are obtained for $\nbf =1,2,4,8$, respectively.\\

This is not the end of the story: the roots that are projected in the center of the diagram are the roots of the derivation algebra $Der(V)$ of the Jordan Pair $V=(\jotn, \jobtn)$. $Der(V)$ includes also a Cartan generator associated to the axis on which the Jordan Pair lies, namely $Der(V) = \gon +\cc$.  $\jotn$ and $\jobtn$ are conjugate representations of $\gon$, the conjugation being the involution in the pair.\\
The algebras $\gon$ are $\ad$, $\ad\oplus\ad$, $\aci$, $\es$ for $\nbf =1,2,4,8$, respectively, and they can also be characterized as \textit{reduced structure} algebras
of $\jotn$ : $\gon\cong str_{0}\left( \jotn\right) $ (see
\textit{e.g.} \cite{McCrimmon,McCrimmon2}).\\

 If we jump to the largest algebra $\eo$ we find the following. First we project the roots on the plane of an $\ad$ diagram. Let's denote by $\adc$ this particular $\ad$. We find the diagram of figure \ref{diagram} for $n=8$ (octonionic Jordan Pairs). In the center sits the subalgebra $\gon=\es$, which in turn is represented by a similar diagram for $n=2$, with complex Jordan Pairs, once projected on the plane of an $\ad$ subalgebra of $\es$, that we denote by $\adf$. In the center of this diagram of $\es$, see figure \ref{diagram} for $\nbf =2$, sits a subalgebra $\ad \oplus \ad$, that we denote by $\adgu \oplus \adgd$ . We have that $\es \oplus \cc$ is the derivation algebra of $(\joto, \jobto)$ and $\ad \oplus \ad \oplus \cc$ is the derivation algebra of $(\jotd, \jobtd)$:
 \be
 \begin{aligned}
 \eo &= \adc \oplus 3 \times (\joto, \jobto) \oplus \gzo \\
 &= \adc \oplus 3 \times (\joto, \jobto) \oplus \adf \oplus 3 \times (\jotd, \jobtd) \oplus \gzd \\
 &= \adc \oplus 3 \times (\joto, \jobto) \oplus \adf \oplus 3 \times (\jotd, \jobtd) \oplus \adgu \oplus \adgd
 , \qquad \text{That's it!} \\
 \end{aligned}
 \ee

\begin{figure}[htbp]
\begin{center}
\ingr{0.5}{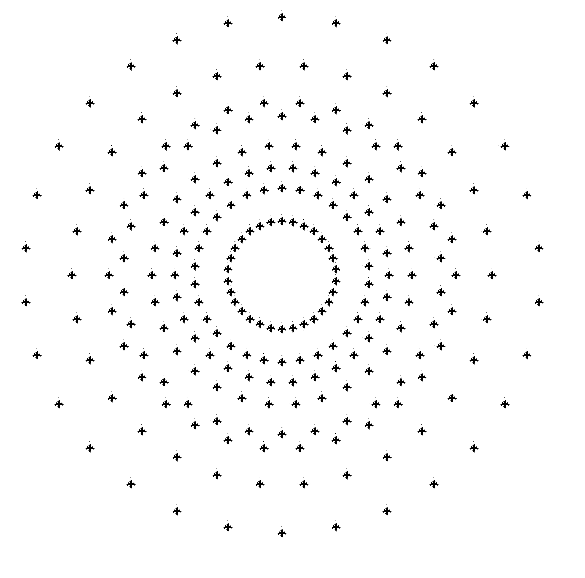}
\end{center}
\caption{The roots of $\eo$ on the Coxeter plane}
\label{cox}
\end{figure}

Since each of the intervening Jordan algebras decomposes into a product of 3-dimensional representations of $\sut$, we have that $\eo$ is actually a clever way of combining 4 $\sut$'s and a bunch of 3-dimensional representations of them \footnote{%
For applications to quantum information theory, \textit{cfr. e.g.} \cite%
{FD-QIT-E6}}. This is not what appears when we view the beautiful projection of the $\eo$ roots on the Coxeter plane, shown in figure \ref{cox}.
As we rotate from the Coxeter plane to the plane of the roots on any of the 4 $\sut$'s we get a diagram like the one shown in figure \ref{diagram}. \\

 The important feature of $\eo$ is that the adjoint representation is the lowest dimensional (fundamental) representation, which means, from the point of you of elementary particles physics, that matter particles and intermediate bosons are all in the same representation.\\

Exceptional Lie groups and algebras appear naturally as gauge symmetries of field theories which are low-energy limits of string
models \cite{Ramond}.

Various non-compact real forms of exceptional algebras occur in
supergravity theories in different dimensions as $U$-dualities\footnote{%
Here $U$-duality is referred to as the \textquotedblleft
continuous\textquotedblright\ symmetries of \cite{CJ-1}. Their discrete
versions are the $U$-duality non-perturbative string theory symmetries
introduced in \cite{HT-1}.} . The related symmetric spaces are relevant by themselves for general relativity, because they are Einstein spaces \cite{Helgason}. In supergravity, some of these cosets, namely those  pertaining to the
non-compact real forms, are interpreted as scalar fields of the
associated non-linear sigma model (see \textit{e.g.} \cite{bt1,bt2}, and also \cite{LA08-review} for a review and list of Refs.). Moreover, they can
represent the charge orbits of electromagnetic fluxes of black holes when the Attractor Mechanism \cite{AM-Refs} is studied (\cite{BH-orbits}; for a
comprehensive review, see \textit{e.g.} \cite{small-orbits}), and they also appear as the \textit{moduli spaces} \cite{FM-2} for extremal black hole
attractors; this approach has been recently extended to all kinds of branes in supergravity \cite{brane orbits}. Fascinating group theoretical
structures arise clearly in the description of the Attractor Mechanism for
black holes in the Maxwell-Einstein supergravity , such as the so-called
magic exceptional $\mathcal{N}=2$ supergravity \cite{MESGT} in four
dimensions, which is related to the minimally non-compact real $\mathbf{e}%
_{7(-25)}$ form \cite{Dobrev-2} of $\est$.

The smallest exceptional Lie algebra, $\gd$, occurs for instance
in the deconfinement phase transitions \cite{G2-1}, in random
matrix models \cite{G2-2}, and in matrix models related to $D$-brane physics
\cite{G2-3}; it also finds application to Montecarlo analysis \cite%
{MonteCarlo}.

$\fq$ enters the construction of integrable models on exceptional
Lie groups and of the corresponding coset manifolds. Of particular interest,
from the mathematical point of view, is the coset manifold $\mathfrak{C}%
\mathbb{P}^{2}=F_{4}/Spin(9)$, the octonionic projective plane (see \textit{%
e.g.} \cite{baez}, and Refs. therein). Furthermore, the split real form $%
\mathbf{f}_{4(4)}$ has been recently proposed as the global symmetry of an
exotic ten-dimensional theory in the context of gauge/gravity correspondence
and \textquotedblleft magic pyramids" in \cite{ICL-Magic}.

Starting from the pioneering work of G\"{u}rsey \cite{gursey80,Gursey-2} on
Grand Unified theories (GUTs), exceptional Lie algebras have been related to
the study of the Standard Model (SM), and to the attempts to go beyond it: for example, the
discovery of neutrino oscillations, the fine tuning of the mixing matrices,
the hierarchy problem, the difficulty in including gravity, and so on. The
renormalization flow of the coupling constants suggests the unification of
gauge interactions at energies of the order of $10^{15}$ GeV, which can be
improved  and fine tuned by supersymmetry. In this framework the GUT gauge group $G$ is expected to be simple, to contain the SM gauge group $%
SU(3)_{c}\times \ SU(2)_{L}\times \ U(1)_{Y}$ and also
to predict the correct spectra after spontaneous symmetry breaking. The
particular structure of the neutrino mixing matrix has led to the proposal
of $G$ given by the semi-direct product between the exceptional group $E_{6}$
and the discrete group $S_{4}$ \cite{E6-GUT}. For some mathematical studies on various real forms of $\es$, see e.g. \cite{Dobrev}, and Refs. therein.

Recently, $\est$ and \textquotedblleft groups of type $E_{7}$"
\cite{brown} have appeared in several indirectly related contexts. They have
been investigated in relation to minimal coupling of vectors and scalars in
cosmology and supergravity \cite{E7-cosmo}. They have been considered as
gauge and global symmetries in the so-called \textit{Freudenthal gauge theory} \cite%
{FGT}. Another application is in the context of entanglement in quantum
information theory; this is actually related to its application to black
holes via the \textit{black-hole/qubit correspondence} (see \cite{BH-qubit-Refs} for
reviews and list of Refs.). For various studies on the split real form of $\est$ and its application to maximal supergravity, see e.g. \cite{KS,KK,Bianchi-Ferrara,Brink}.

The largest finite-dimensional exceptional Lie algebra, namely $\eo$, appears in maximal supergravity \cite{MS} in its maximally non-compact
(split) real form, whereas the compact real form appears in heterotic string
theory \cite{GHMR}. Rather surprisingly, in recent times the popular press
has been dealing with $\eo$ more than once. Firstly, the
computation of the Kazhdan-Lusztig-Vogan polynomials \cite{V} involved the
split real form of $\eo$. Then, attempts at formulating a
\textquotedblleft theory of everything" were considered in \cite{L}, but
they were proved to be unsuccessful (\textit{cfr. e.g.} \cite{DG10}). More
interestingly, the compact real form of $\eo$ appears in the
context of the cobalt niobate ($CoNb_{2}O_{6}$) experiment, making this the
first actual experiment to detect a phenomenon that could be modeled using $%
\mathbf{e}_{8}$ \cite{E8-exp}.

It should also be recalled that alternative approaches to quantum gravity,
such as loop quantum gravity, \cite{carlo} have also led towards the
exceptional algebras, and $\eo$ in particular (see \textit{e.g.}
\cite{smoli}).

There is a wide consensus in both mathematics and physics on the appeal of
the largest exceptional Lie algebra $\eo$, considered by many
beautiful in spite of its complexity (for an explicit realization of its octic invariant, see \cite{Octic}).\\

A nice and practical way of representing the octonions is through the Zorn matrices. We have succeeded  in casting the adjoint representation of all exceptional groups in a Zorn-type matrix form, \cite{T-2}. This has given us a practical way of performing calculations with $\eo$ and apply them to the program we will sketch in the second part of the talk.\\

\section{Part 1. $\mathbf {E_8}$ and Jordan Pair triples}
\nin {\it Nomina sunt consequentia rerum.} Flavius Iustinianus\\

All Lie and Hurwitz algebras in Part 1 are over the complex field $\cc$.\\

\subsection{Jordan Pairs}\label{sec:jp}
In this section we review the concept of a Jordan Pair, \cite{loos1} (see also \cite{McCrimmon,McCrimmon2} for enlightening overviews). Its first appearance in a physics paper, as a quite peculiar quantum mechanical model, is due to Biedenharn, \cite{bied}.\\

Jordan Algebras have traveled a long journey, since their appearance in the 30's \cite{jvw}. The modern formulation \cite{jacob1} involves a quadratic map $U_x y$ (like $xyx$ for associative algebras) instead of the original symmetric product $x \jdot y = \frac12(xy + yx)$. The quadratic map and its linearization $V_{x,y} z = (U_{x+z} - U_x - U_z)y$ (like $xyz+zyx$ in the associative case) reveal  the mathematical structure of Jordan Algebras much more clearly, through the notion of inverse, inner ideal, generic norm, \textit{etc}. The axioms are:
\begin{equation}
U_1 = Id \quad , \qquad
U_x V_{y,x} = V_{x,y} U_x \quad  , \qquad
U_{U_x y} = U_x U_y U_x
\label{qja}
\end{equation}
The quadratic formulation led to the concept of Jordan Triple systems \cite{myb}, an example of which is a pair of modules represented by rectangular matrices. There is no way of multiplying two matrices $x$ and $y$ , say $n\times m$ and $m\times n$ respectively, by means of a bilinear product. But one can do it using a product like $xyx$, quadratic in $x$ and linear in $y$. Notice that, like in the case of rectangular matrices, there needs not be a unity in these structures. The axioms are in this case:
\begin{equation}
U_x V_{y,x} = V_{x,y} U_x \quad  , \qquad
V_{U_x y , y} = V_{x , U_y x} \quad , \qquad
U_{U_x y} = U_x U_y U_x
\label{jts}
\end{equation}

Finally, a Jordan Pair is defined just as a pair of modules $(V^+, V^-)$ acting on each other (but not on themselves) like a Jordan Triple:
\begin{equation}\begin{array}{ll}
U_{x^\sigma} V_{y^{-\sigma},x^\sigma} &= V_{x^\sigma,y^{-\sigma}} U_{x^\sigma}
\\
V_{U_{x^\sigma} y^{-\sigma} , y^{-\sigma}} &= V_{x ^\sigma, U_{y^{-\sigma}} x^\sigma} \\
U_{U_{x^\sigma} y^{-\sigma}} &= U_{x^\sigma} U_{y^{-\sigma}} U_{x^\sigma}\end{array}
\label{jp}
\end{equation}
where $\sigma = \pm$ and $x^\sigma \in V^{+\sigma} \, , \; y^{-\sigma} \in V^{-\sigma}$.

Jordan pairs have two relevant features, both important in the applications to physics and quantum mechanics in particular: they incorporate Jordan algebras and their {\it isotopes}, \cite{McCrimmon2}, in a single structure and they provide a notion of {\it ideal}, the extension of the inner ideals in the quadratic formulation of Jordan algebras.\\

Jordan pairs are strongly related to the Tits-Kantor-Koecher construction of Lie Algebras $\lk$ \cite{tits1}-\nocite{kantor1}\cite{koecher1} (see also the interesting relation to Hopf algebras, \cite{faulk}):
\begin{equation}
\lk = J \oplus \str(J) \oplus \bar{J} \label{tkk}
\end{equation}
where $J$ is a Jordan algebra and $\str(J)= L(J) \oplus Der(J)$ is the structure algebra of $J$ \cite{McCrimmon}; $L(x)$ is the left multiplication in $J$: $L(x) y = x \jdot y$ and $Der(J) = [L(J), L(J)]$ is the algebra of derivations of $J$ (the algebra of the automorphism group of $J$) \cite{schafer1,schafer2}.

 In the case of complex exceptional Lie algebras, this construction applies to $\est$, with $J = \joto$, the 27-dimensional exceptional Jordan algebra of $3 \times 3$ Hermitian matrices over the complex octonions, and $\str(J) = \es \otimes \cc$. The algebra $\es$ is called the \emph{reduced structure algebra} of $J$, $\str_0(J)$, namely the structure algebra with the generator corresponding to the multiplication by a complex number taken away: $\es = L(J_0) \oplus Der(J)$, with $J_0$ denoting the traceless elements of $J$.

We conclude this introductory section with some standard definitions and identities in the theory of Jordan algebras and Jordan pairs, with particular reference to $\jotn, \mathbf{n}= 1,2,4,8$.
If $x,y \in \jotn$ and $xy$ denotes their standard matrix product, we denote by $x\jdot y := \frac12 (xy + yx)$ the Jordan product of $x$ and $y$. The Jordan identity is the power associativity with respect to this product:
\be\label{pass}
x^2 \jdot (x\jdot z) - x \jdot (x^2 \jdot z) = 0,
\ee

Another  fundamental product is the {\it sharp} product $\#$, \cite{McCrimmon}. It is the linearization of $\xs := x^2 - t(x) x - \frac12(t(x^2) - t(x)^2)I$, with $t(x)$ denoting the trace of $x\in \jotn$, in terms of which we may write the fundamental cubic identity for $\jotn, \mathbf{n}= 1,2,4,8$:
\be\label{cubic}
\xs\jdot  x = \frac13 t(\xs\!, x) I \quad \text{or} \quad x^3 - t(x) x^2 + t(\xs) x - \frac13 t(\xs\! , x) I = 0
\ee
where we use the notation $t(x,y) := t(x\jdot y)$ and  $x^3 = x^2 \jdot x$ (notice that for $\joto$, because of non-associativity, $x^2 x \ne x x^2$ in general).

The triple product is defined by, \cite{McCrimmon}:
\bea{ll}\label{vid}
\{ x , y , z \} := V_{x,y}z :&= t(x,y) z + t(z,y) x - (x \# z) \# y \\
&= 2 \left[ (x \jdot y)\jdot z +  (y \jdot z)\jdot x - (z \jdot x)\jdot y \right]
\eea

Notice that the last equality of \eqref{vid} is not trivial at all. $V_{x,y}z$ is the linearization of the quadratic map $U_xy$. The equation (2.3.15) at page 484 of \cite{McCrimmon} shows that:
\be\label{uid}
U_x y = t(x,y) x  - x ^\# \# y = 2 (x \jdot y)\jdot x - x^2 \jdot y
\ee

We shall make use of the following identities, which can be derived from the Jordan Pair axioms, \cite{loos1}:
\be
\left[ V_{x,y} , V_{z,w} \right] = V_{{V_{x,y} z},w} - V_{z,{V_{x,y} w}}
\label{comv}
\ee
and, for $D = (D_+,D_-)$ a derivation of the Jordan Pair $V$ and $\beta(x,y) = (V_{x,y}, - V_{y,x})$,
\be
[D, \beta(x,y)] = \beta (D_+(x),y) + \beta(x, D_-(y))
\label{dib}
\ee


\subsection{The Freudenthal-Tits Magic Square}

\begin{figure}[htbp]
\begin{center}
\ingr{0.6}{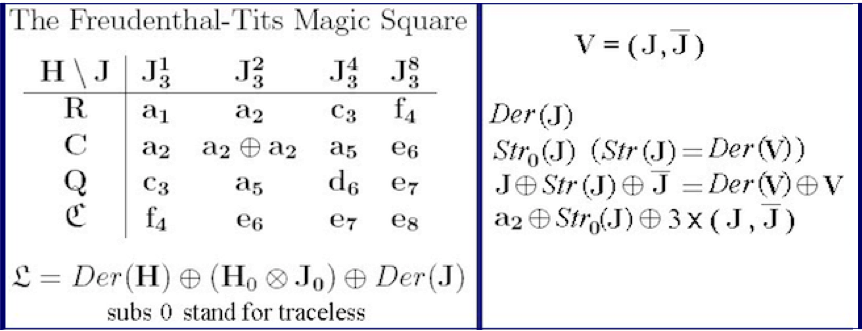}
\end{center}
\caption{The FTMS and its relation to Jordan Pairs}
\label{ftms}
\end{figure}
Two mathematical developments are fundamental in the theory of exceptional Lie algebras. First, the Tits' construction \cite{tits2} which links the exceptional Lie algebras to the four Hurwitz algebras and to the exceptional Jordan algebra of $(3\times 3)$ Hermitian matrices based on the octonions:
\be
\lk = Der(\oo) \oplus (\oo_0 \otimes \jo_0) \oplus Der(\jo)
\ee
where $\oo_0$ and $\jo_0$ denote traceless $\oo$ and $\jo$ respectively.\\
 Second, the Freudenthal-Tits Magic Square (FTMS), \cite{tits2} \cite{freu1}, which is a table of Lie algebras built  with the Tits' construction, but containing also classical algebras, see figure \ref{ftms}.\\

\begin{figure}[htbp]
\begin{center}
\ingr{0.6}{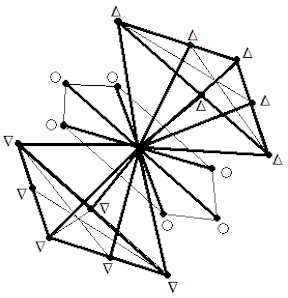}
\end{center}
\caption{Root diagram of $\ctre$: $\ad \oplus \cc$ plus the Jordan Pair $\left( \mathbf{6},\overline{\mathbf{6}}\right) $}
\label{ct_diag}
\end{figure}

The classical algebras of the FTMS can be used as guiding examples for working with the exceptional algebras, except for $\eo$. Think, for instance, of the $\est$ in the third row, fourth column and consider the classical algebra $\ctre$ in the third row, first column. The root diagram of $\ctre$ is shown in figure \ref{ct_diag}. It is a three graded Lie algebra with $\ad$ in degree zero, which is the algebra right above $\ctre$ in the table, and a $\mathbf{6}$ and a $\overline{\mathbf{6}}$ of $\ad$ in degree $\pm 1$. It turns out that the pair $V=\left( \mathbf{6},\overline{\mathbf{6}}\right) $ is a Jordan pair, with triple product $V_{x^{\sigma} ,y^{-\sigma}} z^{\sigma} = [[x^{\sigma},y^{-\sigma}],z^{\sigma}]$, and $Der(V)$ is $\ad \oplus \cc$, where $\cc$ is associated to the Cartan generator along the axis orthogonal to the plane of the $\ad$ diagram.\\
  Figure \ref{diagram} shows that $\fq$ (fourth row, first column) is, in four dimensions,
three copies of $\ctre$ all sharing the same $\ad$ plus 3 Jordan pairs, plus another $\ad$ which {\it rotates} the pairs form one to another. All the planes spanned by the three Jordan pairs $\left( \mathbf{6},\overline{\mathbf{6}}\right) $
are parallel to the plane of the $\ad$ roots and all at the same distance from it.
Notice that in four dimensions there is an infinite number of planes parallel to a given one, all at the
same distance from it.\\

\begin{figure}[htbp]
\begin{center}
\ingr{0.7}{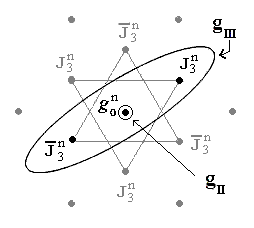}
\end{center}
\caption{The algebras in the IV row of the FTMS, with $\gIII$ and $\gII$ highlighted}
\label{giii}
\end{figure}
The FTMS tells us that the same structure underlies all the algebras of the other columns, just by changing the base ring of the Jordan algebra, as shown in figure \ref{giii}, where $\gII = \gon$ is the algebra of the second row and $\gIII$ that of the third row.\\

In the case of $\eo$ we have that $\gIII = \est$, which is indeed a three graded Lie algebra just like $\ctre$, and $\gII = \es$.

We thus coin a new term: {\em Jordan Pair triple} to mean the structure of 3 Jordan pairs as it appears inside $\eo$ - a triple in the sense of $3\times 1$ and $1\times 3$ rectangular matrices,  in the Zorn-type representation that we show next. This structure is endowed with the extension of the trilinear map  to the triple of Jordan Pairs:
$$V_{x_i^\sigma , y_j^{-\sigma}} z_k^\sigma := [[x_i^\sigma , y_j^{-\sigma}], z_k^\sigma] \ ,\ \sigma = \pm \ , \  i,j,k = 1,2,3$$
Notice that $\eo$ is the smallest simple Lie algebra containing a triple of  $(\joto , \jobto)$ Jordan Pairs plus inner derivations.\\
We paraphrase McCrimmon, \cite{McCrimmon2}, who says that {\it if you open up a Lie algebra and look inside, 9 times out of 10 there is a Jordan algebra (or pair) that makes it tick} and claim that a Jordan pair triple over the octonions is what makes $\eo$ tick.\\

In the next three sections we introduce a Zorn-like matrix representation of $\eo$.


\subsection{Zorn matrix representation of the Octonions}\label{sec:octonions}

\begin{figure}[htbp]
\begin{center}
\ingr{1}{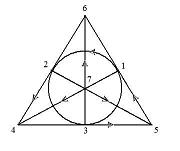}
\caption{Fano diagram for the octonion multiplication\label{fig:fano}}
\end{center}
\end{figure}
In this section we recall the Zorn matrix representation of the octonions. The algebra of the octonions $\oo$ (Cayley numbers) over the complex field  $\cc$  has a multiplication rule that goes according to the Fano diagram in figure \ref{fig:fano} (for earlier studies, see e.g. \cite{Dund}).

If $a \in \oo$ we write $a = a_0 + \sum_{k=1}^7{a_k u_k}$, where $a_k \in \cc$ for $k = 1, \dots , 7$ and $u_k$ for $k = 1, \dots , 7$ denote the octonion imaginary units. We denote by $i$ the the imaginary unit in $\cc$.

We introduce 2 idempotent elements:
$$\rpm = \frac{1}{2}(1 \pm i u_7) $$
and 6 nilpotent elements:
$$\ekpm = \rpm u_k \ , \quad k = 1,2,3 $$
One can readily check that, for $j,k = 1,2,3$:
\begin{equation}
\begin{array}{ll}
 & (\rpm)^2 = \rpm \quad , \quad \rpm \rmp = 0 \\ \\
 & \rpm \ekpm = \ekpm \rmp = \ekpm \quad ,\quad \rmp \ekpm = \ekpm \rpm = 0 \\ \\
 & (\ekpm)^2 = 0 \quad ,\quad \ekpm \ekmp = - \rpm  \quad ,\quad \ejpm \ekmp = 0 \quad j \ne k \\ \\
& \ekpm \varepsilon_{k+1}^\pm = - \varepsilon_{k+1}^\pm \ekpm  = \varepsilon_{k+2}^\mp \qquad \text{(indices modulo 3)} \\ \\
\end{array}\end{equation}

Octonions can be represented by Zorn matrices, \cite{zorn}.
If $a \in\oo$ , $A^\pm \in \cc^3$ is a vector with complex components $\alpha_k^{\pm}$ , $k=1,2,3$ (and we use the
standard summation convention over repeated indices throughout), then we have the identification:
\begin{equation}
a = \alpha_0^+ \rho^+ +\alpha_0^- \rho^- + \alpha_k^+
\varepsilon_k^+ +\alpha_k^- \varepsilon_k^- \longleftrightarrow
\left[
\begin{array}{cc}
\alpha_0^+ &  A^+ \\
A^- & \alpha_0^-
\end{array} \right];
\label{oct} \end{equation}
therefore, through Eq. \eqref{oct}, the product of $a, b \in \mathfrak{C}$ corresponds to:
\begin{equation}
\begin{array}{c}
\left [ \begin{array}{cc} \alpha^+ & A^+ \\ A^- & \alpha^-
\end{array}\right]
\left [ \begin{array}{cc} \beta^+ & B^+ \\ B^- & \beta^-
\end{array}\right] \\ =
\left [\begin{array}{cc} \alpha^+ \beta^+ + A^+ \cdot B^- &
\alpha^+ B^+ + \beta^- A^+ + A^- \wedge B^- \\
\alpha^- B^- + \beta^+ A^- + A^+ \wedge B^+ & \alpha^- \beta^- +
A^- \cdot B^+ \end{array} \right],
\label{zorn1}
\end{array}
\end{equation}
where $A^\pm \cdot B^\mp = - \alpha_K^\pm \beta_k^\mp$ and $A
\wedge B$ is the standard vector product of $A$ and $B$.

\subsection{Matrix representation of $\fq$}\label{sec:f4}

We introduce in this section the representation $\rep$ of $\fq$ in the form of a matrix. For $\fqe \in \fq$:

\begin{equation}
\rep(\fqe) = \left ( \begin{array}{cc} a\otimes I + I \otimes \au & \vxp \\ \vxm & -I\otimes \aut
\end{array}\right),
\label{mfq}
\end{equation}
where $a\in \adu$, $\au \in \add$ (the superscripts being merely used to distinguish the two copies of $\ad$) $\aut$ is the transpose of $\au$,  $I$ is the $3\times 3$ identity matrix, $\vxp \in \cc^3 \otimes \jotu$,  $\vxm \in \cc^3 \otimes \jobtu$ :
$$ \vxp := \left( \begin{array}{c} x_1^+ \\ x_2^+ \\ x_3^+ \end{array} \right) \quad  \vxm := (x_1^- , x_2^- , x_3^- )\ ,  \ x_i^+ \in \jotu \ \ x_i^- \in \jobtu \ , \quad i=1,2,3$$

The commutator is set to be:

\be
\begin{array}{c}
\left[
\left ( \begin{array}{cc} a\otimes I + I \otimes \au & \vxp \\ \vxm & -I\otimes \aut
\end{array}\right) ,
\left ( \begin{array}{cc} b\otimes I + I \otimes \bu &\vyp \\ \vym & -I\otimes \but
\end{array}\right) \right] \\  \\ :=
\left (\begin{array}{cc} C_{11} & C_{12}\\
C_{21} & C_{22}
 \end{array} \right) \hfill
\label{fqcom}
\end{array}
\ee

where:

\be
\begin{array}{ll}
C_{11} &= [a,b] \otimes I + I \otimes [\au,\bu] + \vxp \diamond \vym - \vyp \diamond \vxm \\ \\
C_{12} &=  (a \otimes I) \vyp -  (b \otimes I) \vxp + (I \otimes \au) \vyp + \vyp (I \otimes \aut) \\
 &\phantom{:=} - (I \otimes \bu) \vxp - \vxp (I \otimes \but) +  \vxm \times \vym \\ \\
C_{21} &= - \vym (a \otimes I)  +  \vxm (b \otimes I) - (I \otimes \aut) \vym - \vym (I \otimes \au)  \\
&\phantom{:=} + (I \otimes \but) \vxm + \vxm (I \otimes \bu) +  \vxp \times \vyp \\ \\
C_{22} &=  I \otimes [\aut,\but] + \vxm \bullet \vyp - \vym \bullet \vxp
\end{array}
 \label{comrel}
\ee
with the following definitions :
\be
\begin{array}{ll}
\vxp \diamond \vym &:= \left(\frac13 t(x^+_i, y^-_i) I - t(x^+_i,y^-_j) E_{ij}\right) \otimes I +\\
&\phantom{:=} I \otimes \left(\frac13 t(x^+_i, y^-_i ) I - x^+_i y^-_i \right) \\ \\
\vxm \bullet \vyp &:= I \otimes (\frac13 t(x^-_i,y^+_i) I -  x^-_i y^+_i) \\  \\
(\vxpm \times \vypm)_i &:= \epsilon_{ijk}[x_j^\pm y_k^\pm + y_k^\pm x_j^\pm -x_j^\pm t(y_k^\pm) - y_k^\pm t(x_j^\pm) \\
&\phantom{:=}- (t(x_j^\pm, y_k^\pm) - t(x_j^\pm) t( y_k^\pm)) I]  \\
&:= \epsilon_{ijk} (x_j^\pm \# y_k^\pm)
\end{array}
 \label{not1}
\ee
Notice that:
\begin{enumerate}
\item $x \in \jotu$ is a symmetric complex matrix;
\item writing $\vxp \diamond \vym := c \otimes I + I\otimes \cu$ we have that both $c$ and $\cu$ are traceless hence $c, \cu \in \ad$, and indeed they have $8$ (complex) parameters, and $\vym \bullet \vxp = I\otimes \cut$;
\item terms like $(I \otimes \au) \vyp + \vyp (I \otimes \aut)$ are in $\cc^3\otimes \jotu$, namely they are matrix valued vectors with symmetric matrix elements;
\item the {\it sharp} product $\#$ of $\jotu$ matrices appearing in $\vxpm \times \vypm$ is the fundamental product in the theory of Jordan Algebras, introduced in section \ref{sec:jp}.
\end{enumerate}
In order to prove that $\rep$ is a representation of the Lie algebra $\fq$ we make a comparison with Tits' construction of the fourth row of the magic square, \cite{tits2} \cite{freu1}.
If $\jo_0$ denotes the traceless elements of $\jo$, $\oo_0$ the traceless octonions (the trace being defined by $t(a) := a + \bar a \ \in \cc$, for $a\in \oo$
where the bar denotes the octonion conjugation - that does not affect $\cc$ ), it holds that :
\be
\fq = Der(\oo) \oplus (\oo_0 \otimes \jo_0) \oplus Der(\jo),
\ee

The commutation rules, for $D\in Der(\oo) = \gd $, $c,d \in \oo_0 $, $x,y \in \jo_0$, $E\in Der(\jo)$, are given by:
\be
\begin{array}{l}
\left[ Der(\oo ) , Der(\oo) \right]  = Der(\oo)  \\ \\
\left[ Der(\jo) , Der(\jo)\right]  = Der(\jo)  \\ \\
\left[ Der(\oo), Der(\jo)\right]  = 0  \\ \\
\left[ D, c\otimes x \right] = D(c) \otimes x  \\ \\
\left[ E, c\otimes x \right] = c \otimes E(x)  \\ \\
\left[ c\otimes x, d \otimes y\right] = t(x y) D_{c,d} + 2 (c \ast d)  \otimes (x\ast y)  + \frac12 t(c d)  [x,y]
\end{array}
\label{tc}
\ee

\noindent
where , $\oo_0\ni c\ast d = c d - \frac12 t(c d)$, $\jo_0\ni x\ast y = \frac12 (x y + y x) - \frac13 t(x y) I$, $\jo = \jotu$.

The derivations of $\jo$ are inner: $Der(\jo) = [L(\jo),L(\jo)]$ where $L$ stands for the left (or right) multiplication with respect to the Jordan product: $L_x y = \frac12 (x y + y x)$.
In the case under consideration, the product $x,y \to xy$ is associative and $[L_x,L_y] z = \frac14 [[x,y],z]$. Since $[x,y]$ is antisymmetric, then $Der(\jo) = so(3)_\cc \equiv \aun$.

We can thus put forward the following correspondence:
\be
\begin{array}{ll}
\rep(D) &= \left(\begin{array}{cc} a \otimes I  &  \frac13 tr(\vxp) \otimes I \\ \frac13 tr(\vxm) \otimes I & 0 \end{array}\right) \\ \\
\rep(E)  &= \left(\begin{array}{cc} I \otimes a^A_1  &  0 \\ 0 & I \otimes a^A_1  \end{array}\right)  \\ \\
\rep(\ekp \otimes \jo_0) &= \left(\begin{array}{cc} 0  &  \vxp_k - \frac13 tr(\vxp_k) \otimes I  \\ 0 & 0   \end{array} \right) \\ \\
\rep(\ekm \otimes \jo_0) &= \left(\begin{array}{cc} 0  &  0 \\ \vxm_k - \frac13 tr(\vxm_k) \otimes I  & 0   \end{array} \right) \\ \\
\rep((\rho^+ - \rho^-) \otimes \jo_0) &= \left(\begin{array}{cc} I \otimes a^S_1  &  0 \\ 0 & - I \otimes a^S_1 \end{array} \right),
\end{array}
\label{corr}
\ee
where $a^A_1$ and $a^S_1$ are the antisymmetric and symmetric parts of $a_1$, and $$tr(\vxp) := \left( \begin{array}{c} t(x^+_1) \\  t(x^+_2) \\  t(x^+_3) \end{array} \right) \ , \quad  tr(\vxm) = ( t(x^-_1) ,  t(x^-_2) ,  t(x^-_3))$$
with $\vxpm_k$ denoting a matrix-valued vector whose $k$-th component is the only non-vanishing one.


\subsection{Matrix representation of $\eo$}\label{sec:e8}

Finally, we consider the case of $\eo$, the largest finite-dimensional exceptional Lie algebra.

We use the notation $L_x z := x\jdot z$ and, for $\vx \in \cc^3 \otimes \joto$ with components $(x_1, x_2, x_3)$, $L_\vx \in \cc^3 \otimes L_{\joto}$ denotes the corresponding operator-valued vector with components $(L_{x_1}, L_{x_2}, L_{x_3})$.

We can write an element $\au$ of $\es$ as $\au =  L_x + \sum [L_{x_i},L_{y_i}]$ where $x,x_i,y_i \in \joto$ ($i=1,2,3$) and $t(x) = 0$, \cite{schafer2,jacob2}. The adjoint is defined by $\aud:= L_x - [L_{x_1},L_{x_2}]$. Notice that the operators $F := [L_{x_i},L_{y_i}]$ span the $\fq$ subalgebra of $\es$, namely the derivation algebra of $\joto$  (recall that the Lie algebra of the structure group of $\joto$ is $\es \oplus \cc$).

We should remark that $(\au,-\aud)$ is a derivation in the Jordan Pair $(\joto,\jobto)$, and it is here useful to recall the relationship between the structure group of a Jordan algebra $J$ and the automorphism group of a Jordan Pair $V = (J,J)$ goes as follows \cite{loos1}: if $g \in Str(J)$ then $(g, U^{-1}_{g(I)} g) \in Aut(V)$. In our case, for $g = 1 + \epsilon (L_x + F)$, at first order in $\epsilon$ we get (namely, in the tangent space of the corresponding group manifold) $ U^{-1}_{g(I)} g = 1 + \epsilon (- L_x + F) +O(\epsilon^2)$.

Next, we introduce a product $\mep$,\cite{bt2}, such that $L_x \mep L_y :=  L_{x\cdot y} + [L_x, L_y]$, $F\mep L_x := 2 F L_x$ and $L_x \mep F :=2  L_x F$ for $x,y \in \joto$, including each component $x$ of $\vx \in \cc^3\otimes \joto$ and $y$ of $\vy \in \cc^3\otimes \joto$. By denoting with $[\ ;\ ]$ the commutator with respect to the $\mep$ product, we also require that $[F_1 ; F_2] := 2 [F_1,F_2]$. One thus obtains that $L_x \mep L_y + L_y \mep L_x = 2 L_{x\cdot y}$ and $[F; L_x] := F\mep L_x - L_x \mep F= 2 [F, L_x] = 2 L_{F(x)}$, where he last equality holds because $F$ is a derivation in $\joto$.\\

Therefore, for $\fqe \in \eo$, we write:
\begin{equation}
\rep(\fqe) = \left ( \begin{array}{cc} a\otimes Id + I \otimes \au & \Lvxp \\ \Lvxm & -I\otimes \aud
\end{array}\right)
\label{meo}
\end{equation}
where $a\in \adc$, $\au \in \es$, and we recall that $I$ is the $3\times 3$ identity matrix, as above; furthermore, $Id := L_I$ is the identity operator in $L_{\joto}$ (namely, $L_I L_x= L_x$). Notice that $Id$ is the identity also with respect to the $\mep$ product.

By extending the $\mep$ product in an obvious way to the matrix elements \eqref{meo}, one achieves that $(I \otimes \au) \mep \Lvyp + \Lvyp \mep (I \otimes \aud) = 2 L_{(I \otimes \au) \vyp}$ and  $(I \otimes \aud) \mep \Lvym + \Lvym \mep (I \otimes \au) = 2 L_{(I \otimes \aud) \vym}$.

After some algebra, the commutator of two matrices like \eqref{meo} can be computed to read :
\be
\begin{array}{c}
\left[
\left ( \begin{array}{cc} a\otimes Id + I \otimes \au & \Lvxp \\ \Lvxm & -I\otimes \aud
\end{array}\right) ,
\left ( \begin{array}{cc} b\otimes Id + I \otimes \bu &\Lvyp \\ \Lvym & -I\otimes \bud
\end{array}\right) \right] \\  \\ :=
\left (\begin{array}{cc} C_{11} & C_{12}\\
C_{21} & C_{22}
 \end{array} \right) \hfill
\label{eocom}
\end{array}
\ee
where:
\be
\begin{array}{ll}
C_{11} &= [a,b] \otimes Id + 2 I \otimes [\au,\bu] + \Lvxp \diamond \Lvym - \Lvyp \diamond \Lvxm \\ \\
C_{12} &=  (a \otimes Id) \Lvyp -  (b \otimes Id) \Lvxp +2  L_{(I \otimes \au) \vyp}\\
 &\phantom{:=} - 2 L_{(I \otimes \bu) \vxp} +  \Lvxm \times \Lvym \\ \\
C_{21} &= - \Lvym (a \otimes Id)  +  \Lvxm (b \otimes Id) - 2 L_{(I \otimes \aud) \vym} \\
&\phantom{:=} +2  L_{(I \otimes \bud) \vxm} +  \Lvxp \times \Lvyp \\ \\
C_{22} &= 2 I \otimes [\aud,\bud] + \Lvxm \bullet \Lvyp - \Lvym \bullet \Lvxp.
\end{array}
 \label{comreleo}
\ee
It should be stressed that the products occurring in \eqref{comreleo} do differ from those of  \eqref{not1}; namely, they are defined as follows:
\be
\begin{array}{ll}
\Lvxp \diamond \Lvym &:= \left(\frac13 t(x^+_i, y^-_i) I - t(x^+_i,y^-_j) E_{ij}\right) \otimes Id +\\
&\phantom{:=} I \otimes \left(\frac13 t(x^+_i, y^-_i ) Id - L_{x^+_i \cdot y^-_i} - [L_{x^+_i}, L_{y^-_i}] \right) \\ \\
\Lvxm \bullet \Lvyp &:= I \otimes (\frac13 t(x^-_i,y^+_i) Id - L_{x^-_i \cdot y^+_i} - [L_{x^-_i}, L_{y^+_i}]) \\  \\
\Lvxpm \times \Lvypm &:= L_{\vxpm \times \vypm} = L_{\epsilon_{ijk} (x_j^\pm \# y_k^\pm)}.
\end{array}
 \label{not1eo}
\ee

From the properties of the triple product of Jordan algebras, we derive that $ L_{x^+_i \cdot y^-_i} + [L_{x^+_i}, L_{y^-_i}] = \frac12 V_{x^+_i , y^-_i} \in \es\oplus \cc$, see \eqref{vid}. Moreover, one can readily check that $[a_1^\dagger,b_1^\dagger] = - [a_1,b_1]^\dagger$, $(a\otimes Id) L_b = L_{(a\otimes Id) b}$ and $\Lvym \bullet \Lvxp = I \otimes (\dfrac13 t(x^+_i,y^-_i) Id - L_{x^+_i \cdot y^-_i} - [L_{x^+_i}, L_{y^-_i}])^\dagger$; this result implies that we are actually considering an algebra.

In \cite{T-2} we have proven Jacobi's identity for the algebra of Zorn-type matrices \eqref{meo}, with Lie product given by \eqref{eocom} - \eqref{not1eo}. Once Jacobi's identity is proven, the fact that the Lie algebra so represented is $\eo$ is made obvious by a comparison with the root diagram in figure \ref{diagram}, for $n=8$; in this case, we have:
\begin{itemize}
\item[1)] an $g_{0}^{8}=\es$, commuting with $\adc$;
\item[2)] As in general, the three Jordan Pairs which globally transform as a $(\mathbf{3},\overline{\mathbf{3}})$ of $\adc$; in this case, each of them transforms as a $(\mathbf{27}, \overline{\mathbf{27}})$ of $\es$.
\end{itemize}

As a consequence, we reproduce the well known branching rule of the adjoint of $\eo$ with respect to its maximal and non-symmetric subalgebra $\adc \oplus \es $:
\begin{equation}
\mathbf{248}=\left( \mathbf{8},\mathbf{1}\right) +\left( \mathbf{1},\mathbf{%
78}\right) +\left( \mathbf{3},\mathbf{27}\right) +\left( \overline{\mathbf{3}%
},\overline{\mathbf{27}}\right) .
\end{equation}



\section{Part 2. Fundamental Physics}
\nin {\it Pluralitas non est ponenda sine necessitate.} William of Ockham\\

This second part is about unpublished speculations based on what the first part may suggest or at least suggested us. We will not present here a theory, nor a model, but rather a framework or a scenario which may hopefully attract some attention and stimulate critical thoughts on fundamental physics. The ideas presented here are independent of the structure presented in the first part, that we see as a possible application.\\

We live in a very exciting time for fundamental physics.
On the one hand LHC is producing a large amount of results of observations at smaller and smaller distances; on the other hand many experiments and telescopes on satellites are giving us a deeper and deeper understanding of the universe at the large scale. Being the very small and the very large strictly related in physics, we may say that we have made far-reaching changes in our knowledge of the Universe in the last few decades. This does not mean that we have a better understanding. Many things are far from being explained at a fundamental level and this is part of the present excitement among physicists.\\

We have found a Higgs boson. We hope to find new matter particles with the new run at LHC, at the fantastic energy it has already proven it can work. It will give us some answer on our quest for dark matter and supersymmetric partners of the particles we know.\\
We can look into the past and reconstruct the history of our Universe back to when it had just come out of its state of pure plasma and take a {\it picture} of it at that time, 13.7 billion years ago. We can observe supernovas wherever we point our most advanced telescopes, measure with accuracy the distance of extremely remote galaxies and confirm the accelerating expansion of the Universe.\\
We can make direct observations and experiments confirming with great accuracy quantum mechanics and general relativity, which were unimaginable by their founders.\\

And yet the questions fundamental physics is facing are so crucial that they lead to reconsider both theories and the concept of spacetime itself.
In the long search for simplicity (theory of everything) a conceptual difficulty arises in unifying gravity with the other interactions.
The accelerating expansion of the Universe and the observed motion of the galaxies have forced us to introduce new misterious ingredients: dark energy and dark matter, carrying most of the energy in the Universe.
The incompatibility of quantum mechanics and general relativity at the Planck scale has led to review the concept of spacetime.\\

May a new theory arise from which both quantum mechanics and general relativity derive, that can solve this puzzle and avoid the introduction of dark energy or even dark matter?\\

\subsection{The structure of spacetime}

\nin {\it The creative principle [of science] resides in mathematics.} Albert Einstein\\

The geometrical structure of space has always attracted the attention of mathematicians and physicists at least starting from the work of Gauss. Since then it has been the big open
problem in fundamental physics and has given rise to several conjectures. One of them is
 that spacetime geometry could be based on a p-adic or even a finite field. Beltrametti and Cassinelli in the late 1960's and early 1970's were among the earliest in exploring this idea, \cite{bel}. A major turn in this line of thought came in 1987, when Igor Volovich formulated  the hypothesis that the spacetime geometry may be non archimedean and p-adic at the Planck scale, \cite{vol,vol2}. Varadarajan wrote several important papers on p-adic physics and studied non archimedean models of spacetime, \cite{raja}.\\

The incompatibility of gravity and quantum mechanics at small scales makes it impossible to test if spacetime is a continuum: as you approach the Planck scale you destroy the geometry you were trying to probe and form a black-hole. Spacetime continuum can only be assumed as a possible working hypothesis.\\

Various attempts of going beyond the classical concept of spacetime, in the search for the unification of gravity and quantum mechanics, are currently being made.\\
The theory of strings is the most successful example. Many string theorists believe that spacetime is an emerging concept through the {\it holographic principle}, (see \textit{e.g.} \cite{bou}).\\
{\it Loop quantum gravity}, \cite{carlo}, also describes a non classical spacetime. In particular space is discrete, as a direct consequence of quantization.\\
A quantum spacetime is also considered by many authors who link noncommutative geometry, \cite{connes}, to quantum groups symmetries, \cite{maj}.\\

It seems to us that all these theories start with, or imply, structures somehow related to spacetime.\\

We do believe, following the Ockham's principle, that a genuine new theory should be based on the least possible assumptions on the basic laws of physics and should then push them to their limit.
		
\subsection{No spacetime}

\nin {\it It is my opinion that everything must be based on a simple idea. And it is my opinion that this idea, once we have finally discovered it,
	will be so compelling, so beautiful, that we will say to one another, yes, how could it have been any different.}   J.A. Wheeler\\

An important feature of spacetime is that it is {\em dynamical} and related to matter, as Einstein taught us in his theory of general relativity.\\
The Big Bang, for instance, is not a blast in empty space. Physicists do not think there was a space and the Big Bang happened in it: Big Bang was a blast of spacetime itself, as well as matter and energy. Spacetime stretches like the surface of a balloon that is being blown up, thus explaining why the farther the galaxies are the faster they move away from us. If two galaxies withdraw from each other it simply means that space is being created between them. There is only one way physicists know to reproduce the process of creation: make particles interact.\\

Therefore from the idea that spacetime is dynamical and that it is being created we conclude that:\\

\nin {\it Basic Principle:} There is no way of defining spacetime without a {\em prelimimary} concept of interaction.\\

Stated differently, a universe of non-interacting particles has no spacetime: there is no physical quantity that can relate one particle to another.\\

Our basic principle implies that we have to start with a model of interactions, consistent with the present observations, and deduce from  it what spacetime is.\\

We are used to start from spacetime because our point of view is that of an observer (who measures things in spacetime). However, if we want to describe extreme situations like at the Planck scale or right after the Big Bang, we can no longer give any meaning to the concept of observer. Actually no observation can be made of the Universe before it was 380,000 years old: before that age it consisted of a hot opaque plasma.\\
Furthermore, in a theory with no spacetime to start with, the primordial interacting objects cannot be wave functions nor quantum fields, we agree with Volovich, \cite{vol}.\\

Many standard physical quantities, structures and theories must be emerging. Is all this too vague and hopeless? We believe it is not, and we want to present some concrete argument in favour of such a drastic approach.\\

We start by noticing that all fundamental interactions look similar at short distances. Their basic structure is very simple: it involves only three guys,
like a product in algebra. The first step in our approach is to define objects and {\it elementary interactions}, with the hypothesis in mind, similar to the Bethe \textit{Ansatz} in the Heisenberg model, that every interaction is made of elementary interactions.\\

\begin{figure}[htbp]
\begin{center}
\ingr{1}{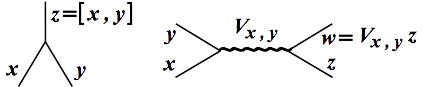}
\end{center}
\caption{Building blocks of the interactions}
\label{int1}
\end{figure}

The proposal is to start from a suitable Lie algebra $\lk$ and state that Lie algebra rules determine the building blocks of interactions. An elementary - or fundamental - interaction may be defined as the interaction between $x$ and $y$ in $\lk$ to produce the outcome $z$ in $\lk$. It is represented by $(x,y ->z) \leftrightarrow [x,y] = z$, see figure \ref{int1}.\\
			
We like to think of $\eo$ as a candidate for $\lk$. Why Lie algebras and $\eo$ in particular? Here are a few possible answers:

\begin{itemize}
\item Current QFT - Yang Mills theory - is based on Lie Algebras;
\item exceptional Lie algebras play a key role in many current theories at various levels;
\item Lie algebras allow to convert geometry into algebra and build a common language between quantum mechanics and gravity where spacetime derivations are commutators and the Leibniz rule $d(xy) = d(x)y+xd(y)$ is the Jacobi identity;
\item the Lie algebra product agrees with the standard quantum mechanical description of a change as a commutator: $[H,G]$ is a change of $G$ by $H$;
\item if $\lk = \eo$ we would have the building blocks of particles and  intermediate bosons in the same adjoint representation. Jordan pairs would perfectly fit into this picture as matter particle-antiparticle pairs, the quarks in the octonionic Jordan pair triple and the leptons in the complex one inside $\es$, while bosons would be represented by the triple product $V_{x_i^\sigma , y_j^{-\sigma}}$.
\end{itemize}

This gives us the building blocks and a rule for the interactions. But where is the geometry hidden and how can interactions give spacetime birth and make it grow?\\

Let us start by defining a point. Thinking of the standard way we define {\it locality}, we reverse its definition and state:\\

\nin {\it Definition of point:} a point is where an interaction occurs.\\

With this in mind we can try to build a toy model for the Big Bang: we start with just the Jordan pair triples of $\eo$, taking primitive idempotents,\cite{bied}, as basis elements ({\it matter particle building blocks}), then fire up the interaction by letting all of them act upon each other by means of the trilinear map $V$, representing the intermediate bosons. The fact that all matter particles interact is interpreted, through our definition of point, by the fact that they all are initially at the same point, even though there is no geometry, no singularity and actually no point of an \textit{a priori} space.\\
This is the first stage (or level) of interactions. We see the outcome and pass to a second stage and so on. We can intuitively associate a {\it universal discrete (quantum of) time} with each stage of interactions.\\
We immediately realize that we are actually working inside the \textit{universal enveloping algebra} of $\lk$, but we also realize that one rule is missing if we want space to expand from the single initial point. We need to state a rule that tells us when two elements interact. Initially all possible interactions occur since we want them at the same point, but what happens after the first interaction? If all possible interactions still occurred, then all the initial elements would still be at the same point: they would have not {\it moved}, or better no space would have been created out of that single point.\\

We have put our algebras on a computer and tested the following tentative rule: {\it two elements interact at the second stage if  the structure constants with which they were generated at the first stage are equal in modulus}. This process can be iterated, by keeping track of the coefficients at each stage. We have tried it with $\es$ at the moment and checked that our toy model of a primordial Universe does ignite: it grows, in number of particles, and expands, since particles interact at different points, created by the previous interactions. The expansion is extremely fast initially, then it slows down. We are currently testing our computer program and analyzing the data it provides. We still are at a very preliminary stage.\\

At this level of our research we are far from determining the structure of spacetime. We only have a few hints, like defining a distance in terms of the number of interactions in a lattice configuration with periodic boundary conditions. It is obvious, however, that the spacetime emerging in our approach is dynamical, finite and discrete, being the outcome of a countable number of interactions among a finite number of objects. This is in agreement with the two cutoffs coming from our current knowledge of physics: the background radiation temperature (finiteness) and Planck length (discreteness). The granularity of spacetime implies that the velocity of propagation of the interaction is also discrete and finite. If the distance traveled from one level of interactions and the next one is 1 Planck length and the time interval is 1 Planck time then the maximum speed of propagation is the speed of light. The model is intrinsically relativistic. It is also obviously quantum mechanical since both the building blocks (which fulfill a quantum logic, \cite{bied}) and their interactions are.\\
We also have that all other infinities or continuities of the standard theories are not present: we have no symmetry Lie groups, just Lie algebras, no transcendental functions, just polynomial functions, by which the \textit{universal enveloping algebra} may be represented by the Poincar\'e-Birkhoff-Witt theorem.\\
Our approach will lead to a finite model by construction, with the continuum limit as a macroscopic approximation, with no extra dimensions, \textit{a priori},  actually with no \textit{a priori} dimension at all. The concepts of wave function and quantum field will be continuum limit approximations of the expansion of our primordial particles as spacetime grows. Polarizations in our approach are also emerging concepts, inherited from the Lie algebra elements, which we assume endowed with a formal right or left handedness, responsible for our perception of the dimensions of space.\\

\section{Conclusion}

\nin {\it If I have communicated even a fraction of the excitement that I feel in thinking about these ideas, I have fulfilled the task I set out for myself.} V.S. Varadarajan, \cite{raja2}\\

The aim of this paper was to give an indication on the possibility of building a model of strongly interacting elementary particles in which spacetime is an emergent concept. In our opinion it involves beautiful mathematics and simple intuitive ideas.

Beauty and simplicity have always guided the research in the history of physics, together with an obstinate effort towards unification.

It has always occurred that the more unified things get, the more undetermined they become, like energy and mass, wave and particle, space and time.\\

\section*{Acknowledgments}
\nin We thank the organizing and scientific committee of the Conference for allowing us to be with Raja for the celebrations in the occasion of his retirement.\\

\nin The work of PT is supported in part by the \textit{Istituto Nazionale di Fisica Nucleare} grant In. Spec. GE 41.\\

\nin Our deepest gratitude goes to Raja, for being so patient in teaching us the right way of thinking.


\end{document}